\documentclass[9pt,twocolumn,twoside]{osajnl}

\journal{jocn} 
\usepackage{fancyhdr}

\setboolean{shortarticle}{false}
\title{Heterogeneous Transmission of Analog Radio and Digital Coherent Signals Over Multi-Span Metro and PON for Bandwidth-Efficient Fronthaul in mmWave Centralized RAN Networks [Invited]}

\author[1]{Devika Dass}
\author[2]{Dan Kilper}
\author[3]{Liam Barry}
\author[1]{Marco Ruffini}

\affil[1]{CONNECT, School of Computer Science and Statistics, Trinity College Dublin}
\affil[2]{CONNECT, School of Engineering, Trinity College Dublin,}
\affil[3]{School of Electronic Engineering, Dublin City University}

\affil[*]{dassd@tcd.ie}

  \dates{ 9 December 2024}

\begin{abstract}
We experimentally investigate the transparent coexistence of heterogeneous Analog Radio-over-Fiber (ARoF) and Digital Coherent Optical (DCO) signals in a converged metro/PON network. Our streamlined setup employs RF generation via optical heterodyning, so that both carrier and modulated signals can be generated centrally and transmitted to the antenna site, across a metro network and Passive Optical Network (PON). The experiment includes the transmission of 8.8 Gbps mmWave signals and 400 Gbps coherent optical signals within a 68.75 GHz ROADM channel bandwidth and a 1:32 to 1:128 split PON. We also analyze the impact of varying ROADM channel bandwidth, PON split ratios, metro network distance and number of ROADMs traversed, on the performance of DCO and ARoF signals. The results reveal that the error vector magnitude (EVM) of the ARoF signal is significantly influenced by the allocated bandwidth, the number of ROADMs, and the overall network loss, providing insights into optimizations necessary to achieve target EVM levels.
\end{abstract}

\begin{document}

\maketitle
% Set footer
\pagestyle{fancy}
\fancyhf{}
\fancyfoot[C]{\footnotesize This paper is a preprint of a paper accepted to JOCN 2025.}

\section{Introduction}

6G mobile transport will support a diverse range of applications, including IoT, cloud computing and AI, requiring significantly higher capacity, peak data rates, and lower latency than its 5G predecessor. 
The emergence of more demanding requirements has driven research to generate renewed interest in greater convergence of optical and wireless networks to enhance the viability of proposed solutions \cite{patwary2020potential}. 
In addition, wireless communications is expected to increase the use of higher frequencies, to increase the bandwidth that can be made available for the next generation of 6G systems. The growing number of human and machine connections within access networks has also accelerated cell site densification, aligning with the short-range characteristics of high-frequency transmission.

%---centralization
The Centralized Radio Access Network (C-RAN) concept enables resource sharing amongst several remote RUs. However, C-RAN alone cannot effectively scale to meet the growing capacity and low-latency demands of 6G networks while simultaneously reducing CAPEX and OPEX. This motivates seeking more efficient resource-sharing solutions in the optical infrastructure. Meeting the capacity and scalability demands of future mobile networks in this case is facilitated through promising advances in photonic processing and networking.  
%--------cost
A recent study \cite{fayad20235g} shows how the assessment of the total cost of ownership (TCO) of the network is important for designing optical fronthaul, and selection of fronthaul solutions, like point-to-point, point-to-multipoint, PON and hybrid PON-free space optics (FSO).% could be employed keeping the TCO in mind.

Spectrally efficient data transmission is another aspect that can contribute to improved network scalability.
Recent research highlights the potential of hybrid digital and analog signal transmission as an effective approach to enable efficient network resource sharing \cite{breyne2017comparison, kanta2022demonstration, li2020digital, dahawi2021low,hu2016converged}.  
%---Arof spectral efficiency
The digital transmission of information following the Common Public Radio Interface (CPRI)/Open Base Station Architecture Initiative (OBSAI) protocols over the fronthaul between the baseband unit (BBU) and radio unit (RU) has limited bandwidth efficiency. Enhanced CPRI (eCPRI) was proposed to enable the fronthaul capacity to scale with the baseband wireless data capacity, enabling more efficient bandwidth use and was fully adopted by the Open RAN community. The proposed 7.2 functional split \cite{8479363} however requires locating some part of the digital baseband processing at the antenna site. In addition, the fronthaul rate is still a few times the rate of the mobile cell capacity, which can lead to increased inefficiency as the data rates of the base stations increase. Beyond CPRI, fronthaul alternatives like Analog Radio over Fiber (ARoF) and Sigma Delta over Fiber (SDoF) \cite{breyne2017comparison} are promising to reduce the spectral occupancy of fiber by over an order of magnitude, compared to digital RoF (DRoF). The radio access point is also simplified, requiring only optical amplification and photodetection, RF amplification and antennas. Recent work \cite{kanta2022demonstration} demonstrated the transmission of hybrid CPRI, intermediate frequency over Fiber (IFoF) and baseband over the wavelength selective switch (WSS) based reconfigurable point-to-multipoint fronthaul in a converged fiber-wireless architecture. 

%----convergence
As mobile networks have advanced, Passive Optical Networks (PON) have become a competitive and cost-effective solution for mobile backhaul and fronthaul, particularly in small cells and heterogeneous networks. As a result, upgrading the Optical Distribution Network (ODN) part of PON systems to support both wireline and wireless services has become a viable strategy for operators \cite{li2020digital}. In \cite{dahawi2021low}, the authors describe the sharing of PON infrastructure for mobile networks as a cost-effective approach; however, the remote generation of optical heterodyne carriers by placing the local oscillator at the Optical Network Unit (ONU) or RU could increase network costs. Other work \cite{hu2016converged} realized combined transmission of 120-channel LTE-like IFoF and baseband signals over a PON network but does not employ higher-frequency wireless transmission, which could enhance system capacity. The transmission of hybrid millimeter wave (mmWave) band radio frequency and baseband signals over a 25 km access network was achieved \cite{li2020digital}, but this work did not explore transmission over a long-haul metro/PON architecture.

\begin{figure}[!t]
    \centering
    \includegraphics[width=1\linewidth]{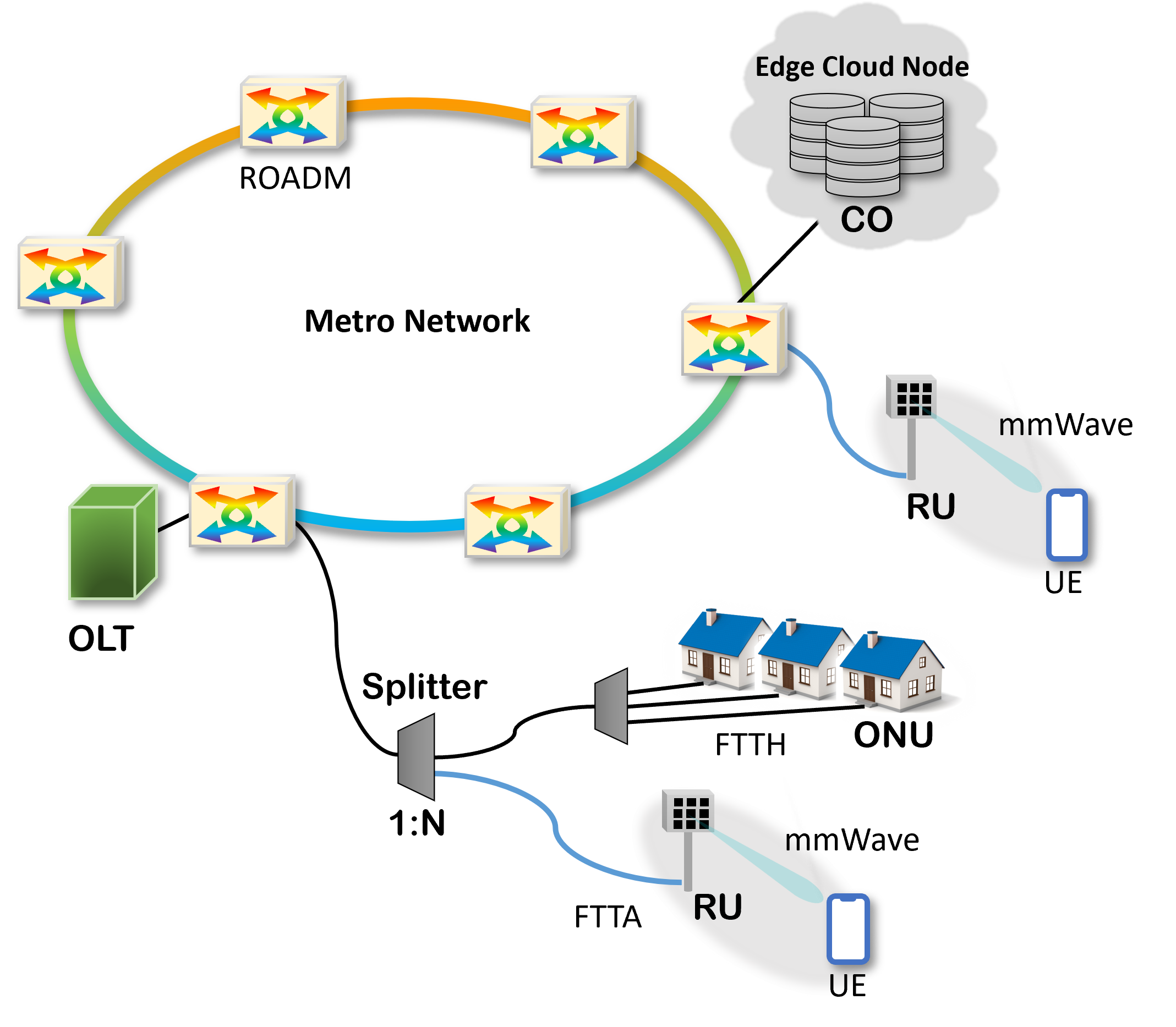}
    \caption{Converged metro-access network, with heterogeneous digital coherent and ARoF transmission. The ARoF signal travels from the edge cloud Central Office node in the metro to the antenna site at a PON endpoint.}
    \label{fig:vision}
\end{figure}

This work aims to analyze the performance of techniques that enable seamless transmission of heterogeneous signals. i.e. Digital Coherent Optical (DCO) and mmWave ARoF across converged metro and PON networks. This is especially beneficial for use cases with high densification of radio access points, at high capacity, which will have to occur at the higher radio frequencies of mmWave and above (where plenty of spectrum is available). To keep costs low in densification scenarios it is important to minimize the cost of the radio access point, moving all digital processing to a central location, for example, an edge node located somewhere in the metro area. 

For this reason, our focus is on the spectrum efficiency of ARoF to distribute radio waveforms from the central office in the metro area to the antenna site, on a PON termination. 
In addition, in such a converged metro-access scenario, the spectrum usage in the metro part of the network becomes the bottleneck, as each metro network will serve many access networks. 
In this work we present the experimental results of this type of scenario, focusing on three key principles:
%- minimize spectrum occupancy of ARoF signals
\begin{enumerate}
    \item Minimise the spectrum occupancy of ARoF signals in the metro network. We achieve this by embedding the radio signals within the allocated spectrum of existing DCO signals. This is to reduce spectrum inefficiency, when the band of the wireless signal is of a similar order of magnitude (or below) than the spectrum allocation granularity of WSSs (typically 6.25 GHz).
    \item Unify the design, linking cloud nodes, central offices and data centers (DC) in metro networks with RUs at PON endpoints, using existing infrastructure as depicted in Fig. \ref{fig:vision}. Seamless data transmission across metro and access networks is achieved using only equipment available today in typical commercial networks. Only the endpoints are introducing new technology, where required. %This is essential to avoid introducing additional costs to the network.
    \item Take advantage of the power split nature of PONs to share spectrum usage. We apply this principle by employing optical heterodyning, where the optical carrier required to generate the RF signal at the endpoints (through optical heterodyning with the ARoF modulated signal) can be shared across all endpoints in the same PON. This technique also simplifies the endpoint, which does not need an RF generator, and it can be cost-effective, especially when moving to even higher frequencies (i.e., sub-THz transmission).

\end{enumerate}

\begin{figure*}[!t]
    \centering
    \includegraphics[width=0.9\linewidth]{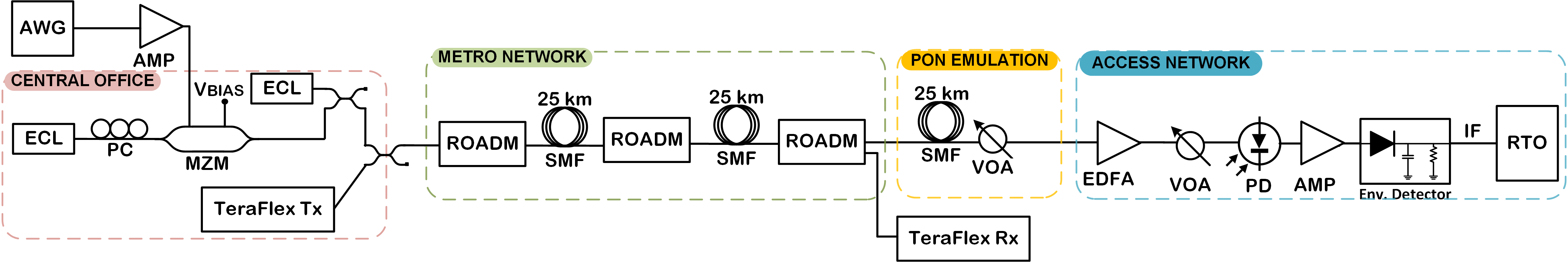}
    \caption{Experimental set up with converged 50 km metro fibre, one in-line and two add drop ROADMs, and 25 Km PON.}
    \label{fig:exp_set_1}
\end{figure*}

\begin{figure*}[!t]
    \centering
    \includegraphics[width=1\linewidth]{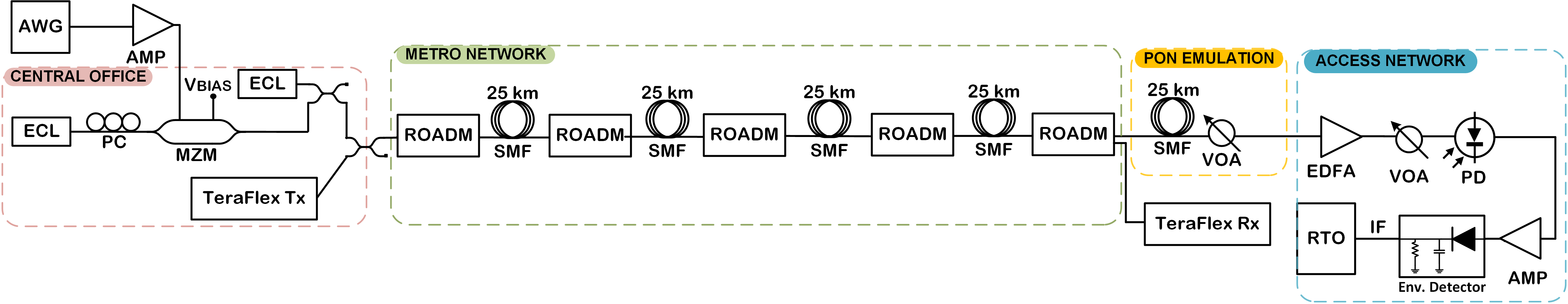}
    \caption{Experimental set up with converged 100 km metro fiber, three in-line and two add drop ROADMs and 25 Km PON.}
    \label{fig:exp_set_2}
\end{figure*}

Previously, we explored the coexistence of DCO and ARoF signals transmitted at an intermediate frequency of 2 GHz, analyzing performance degradation as the signals passed through multiple ROADMs \cite{ICC2023,delmade2023metro}. This work introduces the use of mmWave radio frequency (RF) signals in the 60 GHz band, using optical RF carrier generation. This technique allows placing both modulated signals and secondary carriers remotely at the central office and transmitting them over fiber across the metro and access networks. This technique is also important for scalability, as integrating metro with PON access, allows sharing the optical carrier among multiple mmWave base stations (i.e., exploiting the power split nature of PONs).

Building on prior techniques \cite{dass2022wavelength}, the modulated carrier and secondary carrier beat at the photodetector to produce an RF carrier at a frequency equal to the difference between the two optical tones (55–65 GHz in this work). This approach simplifies the radio head by eliminating the need to generate optical or RF carriers at the RU. To enhance efficiency, we position the ARoF-modulated signal and secondary optical carrier within an existing ROADM channel that already carries a DCO signal. This avoids the inefficiency of creating additional ROADM channels, which require 6.25 GHz granularity and guard bands, for radio signals that might be only a few GHz (or less) in bandwidth. As a result, the DCO signal is spectrally situated between the modulated ARoF signal and the secondary carrier.  

%This paper analyzes the performance of a cost-effective approach where both the modulated signal and carrier are remotely generated and transmitted across the metro and PON access networks. 
To prioritize simplicity and reduce costs, a basic envelope detector for the ARoF signal is employed and the use of independent optical filters is avoided. The DCO signal is filtered out by a wavelength selective switch (WSS) at the egress ROADM in the metro network.
%We further investigate the performance of hybrid signal transmission across the multi-span ROADM network, considering various optical split ratios within the PON network and different channel bandwidths allocated in the metro network.
%Our results show heterogeneous transmission of 400 Gbps DCO signal and at a maximum of 8.8 Gbps 64QAM OFDM ARoF signal at 60GHz in the converged metro/PON network scenario as shown in Fig. \ref{fig:exp_set_1} with a total 75 km fiber and power split of 1:32 in PON ODN, with Q-factor of 8 dB and EVM of 7\% respectively.
This analysis operates over a multi-span ROADM network, with different optical split ratios for the PON network and different channel bandwidths allocated in the metro network.
This work realizes the coexistence of a 400 Gbps DCO signal and a 64QAM OFDM ARoF signal at 60 GHz, achieving a maximum transmission rate of 8.8 Gbps within a converged metro/PON network scenario (illustrated in Fig. \ref{fig:exp_set_1}). This setup involves a total fiber length of 75 km, transmission window size of 68.75 GHz and a power split ratio of 1:32 in the PON ODN, achieving a Q-factor of 8 and an EVM of 7\%, respectively.

%_________________________EXP____________________________________
\section{Experimental Setup}
\label{Exp setup}

\begin{figure}[ht]
    \centering
    \includegraphics[width=0.6\linewidth]{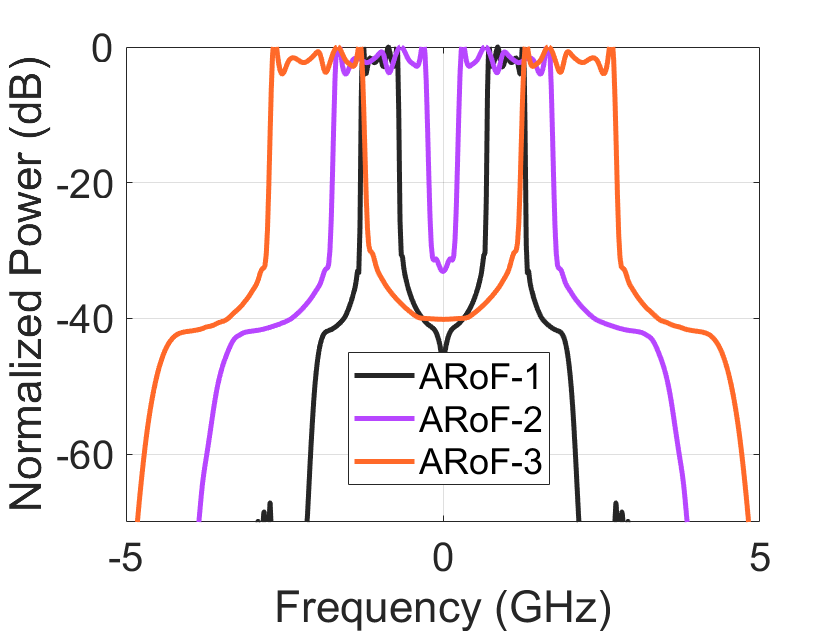}
    \caption{Overlayed electrical spectra of the envelopes of the three ARoF signals, generated in software and utilized in the experiments, provide a visual comparison of their spectral characteristics.}
    \label{fig:AROF overlay}
\end{figure}

\begin{figure}[ht]
    \centering
    \includegraphics[width=0.6\linewidth]{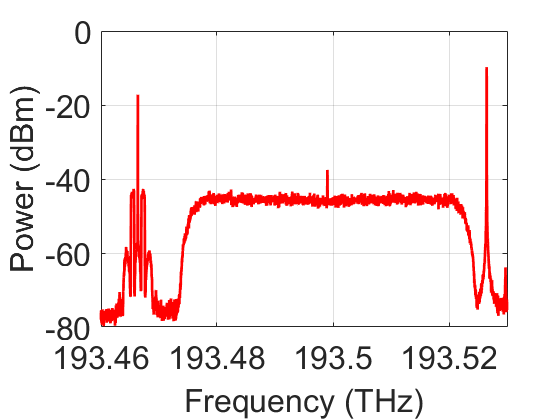}
    \caption{The combined optical spectra of signals launched into the metro network, the modulated ARoF on the left, the DCO in the middle and the optical carrier used for heterodyning on the right hand side.}
    \label{fig:tx spec}
\end{figure}

The two experimental setups, depicted in Fig. \ref{fig:exp_set_1} and \ref{fig:exp_set_2}, combine ARoF signals with varying bandwidths and modulation formats alongside DCO signals with different data rates. It should be noted that we obtain different rates solely by changing the modulation format, keeping the baud rate constant for the DCO signals. This is important as we aim to keep the WSS ROADM bandwidth the same for all DCO rates.  The detailed information about the ARoF and DCO signals is presented in Tab. \ref{tab:sig_parameters}. The total spectral bandwidth occupied by the ARoF signals is also adjusted by changing the intermediate frequencies (IF), which changes the proximity of the two sidebands of the double-sideband (DSB) signal as shown in Fig. \ref{fig:AROF overlay}. The optical spectra of a 3.52 Gbps 16QAM OFDM ARoF signal (ARoF-1) and a 400 Gb/s 64QAM DCO signal are presented in Fig. \ref{fig:tx spec}.

\setlength{\tabcolsep}{2pt}
\renewcommand{\arraystretch}{1.5}
\newcolumntype{s}{>{\columncolor[HTML]{AAACED}} p{cm}}
\begin{table*}[!t]
\centering
\caption{Transmitted Heterogeneous Signal Parameters}
 \label{tab:sig_parameters}

    \begin{tabular}{|c|c|c|c|c|c|c|}
    \hline  %\multicolumn{2}{|c|}
    Parameters            &  ARoF-1 & ARoF-2 & ARoF-3 &  DCO-1     & DCO-2     & DCO-3     \\
    \hline
       
    Bitrate (Gb/s)       & 2.34/3.52    &  5.86/8.79       & 5.86/8.79        & 200       & 300       & 400              \\
    Modulation           &16QAM/64QAM     & 16QAM/64QAM     & 16QAM/64QAM     & DP-P-16QAM  & DP-16QAM  & DP-64QAM          \\
        
    %Ingress WSS Bandwidth (GHz)  & 18.75      &   18.75      &   18.75  &  62.5       & 62.5        &  62.5           \\
    OFDM Bandwidth (GHz)    &  0.6 & 1.5  & 1.5  & N.A.  & N.A.  & N.A.  \\
    Intermediate Frequency (GHz)             & 1        & 1         & 2       & N.A.         & N.A.         & N.A.                \\
        \hline  
    \end{tabular}

\end{table*}

An arbitrary waveform generator (AWG) is used to generate 5G NR analog waveforms with two sidebands at an IF of 1 GHz (for ARoF-1 and ARoF-2) or 2 GHz (for ARoF-3); the signal is amplified using a low-noise electrical amplifier. The waveforms are applied to a dual-drive Mach-Zehnder modulator (DD-MZM) to modulate an optical carrier (15 dBm) emitted by an external cavity laser (ECL). Another ECL provides an optical carrier offset by a value that is between 55 and 65 GHz from the modulated signal, facilitating optical heterodyning at the receiver to produce an mmWave RF signal carrying the 16/64 QAM OFDM 5G waveform within the 55–65 GHz frequency range. %The 55 to 65 GHz value was selected as this is the pass band of the RF amplifier. In our setup, the passband is used as a filter 

The modulated and unmodulated ARoF optical signals are transmitted adjacent to a DCO signal generated by a C-band tunable integrated transceiver (Adtran Teraflex) for transmission through the metro network. This network includes ROADMs (Lumentum), used at the line system's ingress, egress, and intermediate locations. The ROADMs also incorporate erbium-doped fiber amplifiers (EDFAs). ROADMs have booster EDFAs after the add ports and preamp EDFAS after the line in port. In this setup, the booster amplifier of the ingress ROADM is configured with an 8 dB gain, while the preamplifier of the egress ROADM is set to a 20 dB gain. Other amplifiers are adjusted to compensate for preceding losses in the fiber spans and nodes, ensuring a launch power of approximately -2 dBm in the fiber. All components in the experimental setup are interconnected via a reconfigurable fiber space switch (Polatis), which allows us to move between the two target topologies without any manual re-patching.

\begin{figure}[!b]
    \centering
    \includegraphics[width=0.6\linewidth]{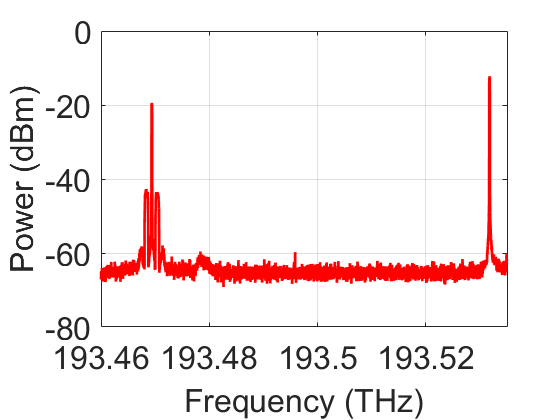}
    \caption{The optical spectrum of the signal received at the ARoF receiver after the egress ROADM dropped the DCO signal; this remaining signal propagates over the PON network.}
    \label{fig:ARoFRX spec}
\end{figure}

The experiment is set up to represent two metro networks: one with 75 km optical fiber as shown in Fig. \ref{fig:exp_set_1} and another with 125 km optical fiber depicted in Fig. \ref{fig:exp_set_2}. At the egress ROADM of the metro network, the DCO and ARoF signals are separated using the WSS filter in the ROADMs and sent to two different drop ports. The optical spectrum of the filtered ARoF signal is shown in Fig. \ref{fig:ARoFRX spec}. One drop section of the egress ROADM sends the DCO signal to a coherent receiver (Adtran Teraflex). The ARoF signal is instead sent to its termination, after traveling through the Optical Distribution Network (ODN) of a PON. %, constituted of 25 km SMF and a variable optical attenuator to emulate power split ratios of 1:32, 1:64 and 1:128.
The PON setup includes a 25 km fiber span and a variable optical attenuator (VOA), emulating the loss of 1:32, 1:64 and 1:128 split ratios, with a total attenuation of 17 dB (15 dB nominal splitter loss plus 2 dB for non-ideal losses-), 19 dB and 21 dB respectively \cite{splitter-loss}. After passing through the metro and PON networks, the mmWave ARoF and DCO signals are amplified at the PON termination using an EDFA to recover the signal from power splitter (PS) losses. %Unlike typical shared broadband services, this PON fiber is dedicated to supporting a 200/300/400 Gb/s service. In practical applications, end users could enhance signal quality by using small ROADMs or cost-effective optical amplifiers.

At the ARoF receiver, a 70 GHz photodetector (PD) converts the signal, at a received optical power $\sim 2 dBm$, to a 5G NR signal at $\sim$60 GHz. The RF signals within the 55–65 GHz range are amplified and sent to an envelope detector, which downconverts the mmWave signal to an IF signal. These samples are recorded using a real-time oscilloscope (RTO) and post-processed for error performance evaluation. No additional filtering is needed, as the RF amplifier’s passband automatically excludes unwanted signals.

\begin{figure*}[ht]
    \centering
    \includegraphics[width=0.7\linewidth]{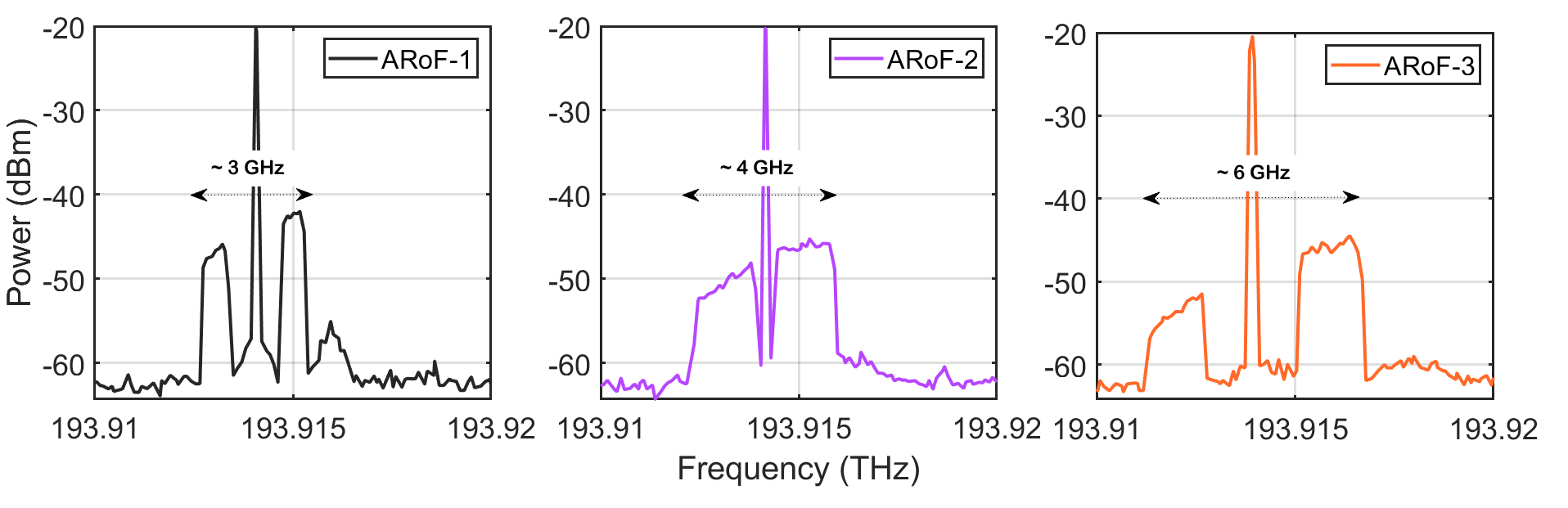}
    \caption{Optical spectra of the different types of ARoF signals at the egress ROADM WSS filter illustrate how the filtering characteristics affect each signal. }
    \label{fig:Optical Spec_ARoF }
\end{figure*}

%--------------R&D-------------------------------------------

\section{Results and Discussion}

\begin{figure*}[!ht]
\centering
\includegraphics[width=1\linewidth]{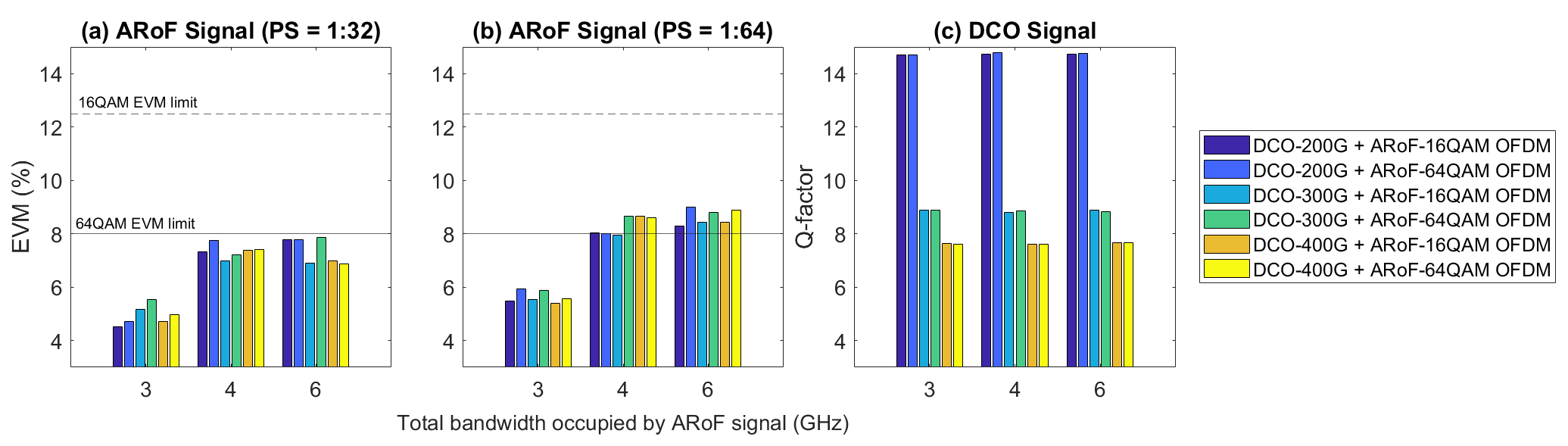}
\caption{EVM vs ARoF DSB bandwidth occupancy for a 75 km metro/PON network with (a) 1:32 and (b) 1:64 split PON emulation. Q-factor vs ARoF DSB bandwidth occupancy for 200G, 300G and 400G DCO signals for the same network. All these are measured with a transmission window size of 68.75 GHz. }
\label{fig:varying DCO datarate}
\end{figure*}

In previous work \cite{ecoc2024}, we analyzed the impact of the transmission window size (62.5 GHz) on the performance of the joint DCO and ARoF signals transmission within the same ROADM channel. This produces an additional loss in the ARoF signal due to the ROADM WSS filter profile, as shown in Fig. \ref{fig:Optical Spec_ARoF }, affecting especially the ARoF-3 signal with the widest spectral bandwidth. 
The properties of the ARoF signals, as defined in Tab. \ref{tab:sig_parameters}, indicate variations in their intermediate frequencies (IF). While ARoF-2 and ARoF-3 share the same OFDM bandwidths, their IFs differ, being 1 GHz and 2 GHz, respectively. As illustrated in Fig. \ref{fig:Optical Spec_ARoF }(b) and (c), ARoF-2's lower IF results in reduced spacing between its sidebands compared to ARoF-3. This configuration enhances spectral efficiency by allowing the DSB signal to occupy less spectrum but increases susceptibility to signal-to-signal beat interference (SSBI).

In this work, we have accounted for WSS bandwidths greater than 62.5 GHz, increasing by multiples of 6.25 GHz, based on WSS granularity. In addition, we provide an extensive analysis over different size metro networks, different PON splits and also use diverse ARoF and DCO signals.

\subsection{DCO and ARoF transmission over 50 km metro and 25 km PON}

In the first experiment, we analyze the impact of varying the DCO signal's datarate on the coexisting ARoF and DCO with 68.75 GHz WSS bandwidth, for the 75 km metro/PON network (see Fig. \ref{fig:exp_set_1}) with PON split ratios of 1:32 and 1:64. The performance indicator for ARoF and DCO signals are EVM (\%) and Q-factor, respectively, for different values of the total bandwidth occupied (in GHz) by the ARoF signal and is depicted in Fig. \ref{fig:varying DCO datarate}. For the 1:32 split ratio, the EVM of the ARoF signals for all the heterogeneous ARoF and DCO signals is below the 8 \% 64QAM FEC threshold (see Fig. \ref{fig:varying DCO datarate} (a)). The value of the Q-factor corresponding to the 200 Gbps, 300 Gbps and 400 Gbps DCO signals is $\sim$ 15, 9 and 8, respectively.

The EVM of the ARoF is larger for the signals with the lowest total spectral bandwidth (ARoF-1 with $\sim$3 GHz). This is attributed to the least impact of ROADM filter penalties and to lower interference with the DCO signal. From the perspective of the DCO Q-factor, no noticeable differences are observed concerning the ARoF bandwidth. This indicates that the DCO signals are resilient to interference caused by the proximity of the ARoF signal. It should also be noted that the DCO signals are dropped at the metro node, so their performance does not depend on the PON split ratio.

In Fig. \ref{fig:varying DCO datarate} (b), the EVM performance of the ARoF signal deteriorates as the PON split ratio is increased to 1:64. The signal SNR is reduced due to the higher loss of the signal before entering the EDFA. 
However, for all the ARoF-1 signals with 3 GHz spectral bandwidth, the EVM remains below the 8 \% 64QAM FEC limit. For the ARoF-2 and ARoF-3 signals, with approximate spectral bandwidths of 4 GHz and 6 GHz respectively, the EVM is at or slightly above the 8\% FEC threshold. However, for all ARoF signals, the EVM is below the 12.5\% 16QAM FEC limit. 

\begin{figure}[!ht]
\centering
\includegraphics[width=1\linewidth]{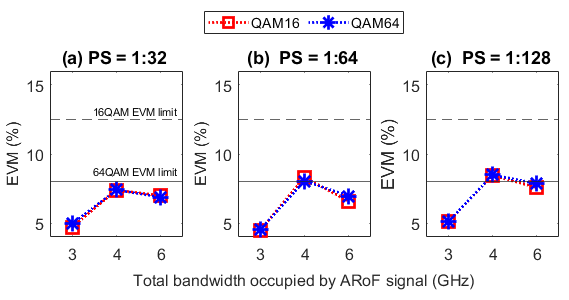}
\caption{A comparison of EVM vs ARoF DSB bandwidth occupancy for a 75 km metro/PON network with (a) 1:32 (b) 1:64 split 
 and (c) 1:128 PON emulation for the transmission window size set to 68.75 GHz. }
\label{fig:vary_split_68}
\end{figure}

Fig. \ref{fig:vary_split_68} (a), (b) and (c), show the impact of increasing the split ratio, 1:32. 1:64 and 1:128, on the EVM of the ARoF signal for the same network scenario. The EVM values for 16QAM and 64QAM coincide for most cases. Here, the worst performance of the ARoF-2 signal observed for all the PS ratios is attributed to SSBI, caused by the smaller spacing between its two sidebands than the bandwidth of each sideband (see Fig. \ref{fig:AROF overlay}). We can thus conclude that in the case of only one in-line ROADM, the SSBI effect due to the lower IF of ARoF-2 is more penalising than the filtering effect of the WSS on the larger bandwidth ARoF-3 signal. As expected, the performance of the ARoF further degrades for the 1:128 split because of lower SNR in this case.

%-------------------WSS=81.25 gHz-------------------------
\begin{figure}[!ht]
\centering
\includegraphics[width=.75\linewidth]{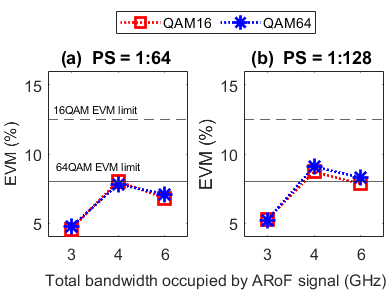}
\caption{A comparison of EVM vs ARoF DSB bandwidth occupancy for a 75 km metro/PON network with (a) 1:64 and (b) 1:128 split PON emulation for WSS bandwidth set to 81.25 GHz. }
\label{fig:vary_split_81}
\end{figure}

\begin{figure}
    \centering
    \includegraphics[width=1\linewidth]{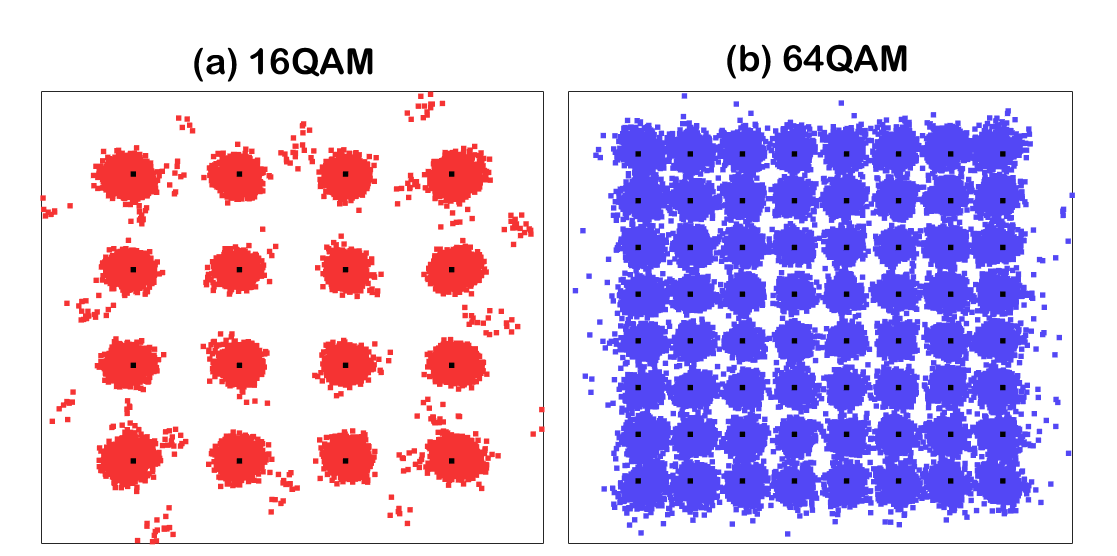}
    \caption{Overlayed constellation of transmitted (shown darker) and received (a) 16QAM and (b) 64QAM ARoF-3 signal over 75 km metro/PON network with 1:64 PON splitter and WSS bandwidth set to 81.25 GHz.}
    \label{fig:const-16qam}
\end{figure}

In Fig. \ref{fig:vary_split_81}, the transmission window size of the ROADMs is increased to 81.25 GHz. The EVM value shows a slight degradation for the ARoF signal in the 1:128 PS ratio scenario (see Fig. \ref{fig:vary_split_81} (b)) compared to the same PS ratio with a 68.75 GHz wide transmission window (refer to Fig. \ref{fig:vary_split_68} (c)). However, the EVM performance remains the same when the PS ratio is 1:64 and the transmission window size is 68.75 GHz and 81.25 GHz (refer Fig. \ref{fig:vary_split_68} (b) and Fig. \ref{fig:vary_split_81} (a)). This implies that for the 75 km metro/PON network scenario, a higher transmission window size (>= 68.75 GHz) doesn't lead to an improvement in error performance. Again the performance of the ARoF-2 signal is worse here due to the SSBI. The overlayed constellations of transmitted and received 16QAM and 64QAM ARoF-3 signals for topology with 1:64 splitter is depicted in Fig. \ref{fig:const-16qam}. 
The above results show the interference of DCO and ARoF signals is negligible even if the two are separated by the default ROADM WSS drop filter, rather than a dedicated fine-tunable filter. %Hence, it can be noted that the variation of the DCO signals data rate didn't impact the performance of ARoF signals and vice versa.

%-----------------------------------------------------

\subsection{DCO and ARoF transmission over 100 km metro and 25 km PON}

\begin{figure}[!t]
\centering
\includegraphics[width=1\linewidth]{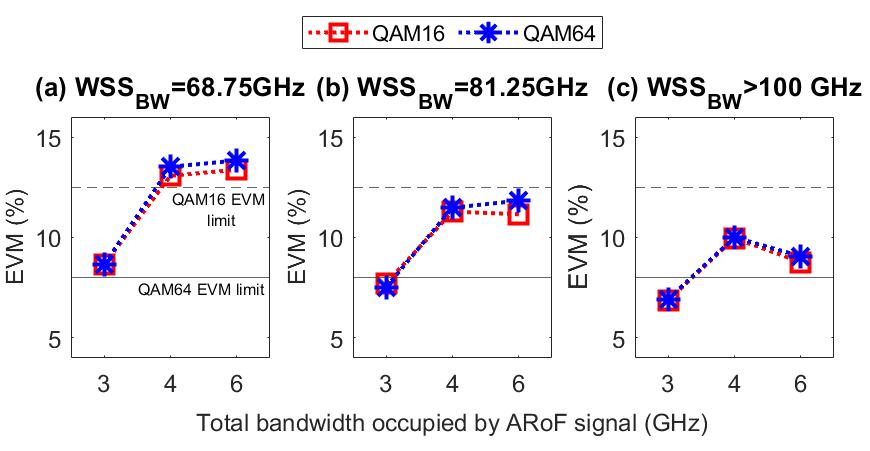}
\caption{EVM vs ARoF DSB bandwidth occupancy for a 125 km metro/PON network with 1:64 split PON emulation.}
\label{fig:4-span 1:64 split}
\end{figure}

In this second experiment, we analyze the the impact of the transmission window size for a metro transmission of 100 km, with four spans, and a 25 km PON fiber, shown in Fig. \ref{fig:exp_set_2}. This experiment is conducted with 1:64 and 1:128 split ratios. Three transmission window sizes are selected in the network scenario: 68.75 GHz, 81.25 GHz and >100 GHz ( i.e. OPEN). Fig. \ref{fig:4-span 1:64 split} shows the EVM results for the ARoF signal with the 1:64 splitter in the PON emulation. In this case, the cascade filtering effect is evident due to the misalignment of center frequencies of the ROADM's WSS filters. When using a WSS filter bandwidth of 68.75 GHz, shown in Fig. \ref{fig:4-span 1:64 split} (a),  all the EVM values lie above the 8\% threshold and only ARoF-1 signal is below the 12.5\% 16QAM FEC limit. Here, the performance of ARoF-3 is worse than that of ARoF-2 because the losses caused by the filtering effect are more dominant than the impact of SSBI.

As the WSS filter bandwidth in the network is increased to 81.25 GHz, the EVM values achieved are lower than the 12.5\% threshold for the ARoF-2 and ARoF-3 signals and touch the 8\% for the ARoF-1 signal. However, the dominant impact of filtering remains evident in Fig. \ref{fig:4-span 1:64 split} (b). In this case, the cascade of the filtering effects of multiple WSSs, due to center frequency misalignment, is worse than the SSBI effect on ARoF-2 (Fig. \ref{fig:4-span 1:64 split} (a) and (b)). This implies that as the number of spans in the metro network increases, we observe the limiting effect of the ROADM WSS filter bandwidth on the performance of the ARoF signals, especially if they lie close to the edge of the WSS transmission window. 
As the WSS bandwidth increases, the filtering effect diminishes, resulting in improved performance of ARoF-3 compared to ARoF-2 (Fig. \ref{fig:4-span 1:64 split} (c)). Notably, the EVM values further improve when the window size exceeds 100 GHz; however, the dominance of the SSBI effect leads to a decline in the EVM performance of the ARoF-2 signal. 

Different types of ARoF signals, as described in Section \ref{Exp setup}, exhibit varying spectral bandwidths, and the results reveal that their performance is strongly influenced by the choice of WSS filter bandwidth. This dependency arises from factors such as the cumulative narrowing effect of multiple filter passes and the drift in center frequencies and passbands, commonly referred to as filter wandering.
Therefore, we can see that there is an interesting trade-off in the choice of the IF value: reducing the IF introduces an SSBI penalty, but reduces the WSS filtering effect. The optimal choice depends on the number of ROADMs traversed by the signal and on the bandwidth that can be allocated to the signal.

%  between reducing the ARoF signal bandwidth at the expense of a higher SSBI, and increasing significant impact of ROADM filtering-induced loss is observed on the widest bandwidth ARoF signal in this longer fiber span network scenario for WSS filter bandwidths of 68.75 GHz and 81.25 GHz. However, as the filter bandwidth is further increased, the effect of SSBI becomes more pronounced, and the ARoF-2 has a worse performance.

%However, when the width of the filter is increased to 81.25 GHz and >100GHz, for only 16QAM in Fig. \ref{fig:4-span 1:64 split} (b) and 16QAM and 64QAM in Fig. \ref{fig:4-span 1:64 split} (c), the ARoF-2 signal has the worse performance because of the SSBI (see Fig. \ref{fig:AROF overlay}). The performance of the 64QAM ARoF signal degrades more significantly than the 16QAM ARoF signal in a 125 km network scenario. This degradation is primarily attributed to reduced SNR caused by the combined filtering effects of multiple ROADMs with slightly misaligned center frequencies, which particularly impacts the wider bandwidth ARoF-3 signal. As a result, the constellation of the 64QAM signal appears more compressed at this SNR compared to the 16QAM signal, reflecting its greater sensitivity to reduced signal quality. 

%----1:128 splitter-------------------------------------------------
\begin{figure}[!t]
\centering
\includegraphics[width=1\linewidth]{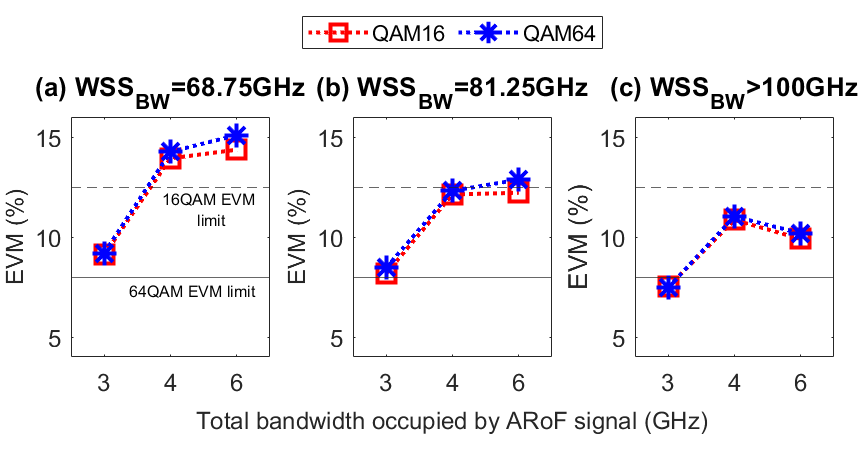}
\caption{EVM vs ARoF DSB bandwidth occupancy for a 125 km metro/PON network with 1:128 split PON emulation.}
\label{fig:4-span 1:128 split}
\end{figure}

Fig. \ref{fig:4-span 1:128 split} shows the performance evaluation for a 1:128 split PON. The increased power loss due to the 1:128 splitter ratio results in a lower SNR, degrading the ARoF signal's EVM compared to the case with a 1:64 splitter ratio. Similar to the results shown for the 1:64 split in Fig. \ref{fig:4-span 1:64 split}, as the filter bandwidth increases, the EVM of the ARoF signal improves, as seen in the previous case. 
For filters with bandwidths greater than 100 GHz, as shown in Fig. \ref{fig:4-span 1:128 split} (c), the EVM of the ARoF-2 signal is worse, primarily due to SSBI caused by the spacing of its two sidebands being less than the bandwidth of each sideband (see Fig. \ref{fig:AROF overlay}). Although closer proximity of the two sidebands leads to higher interference, the DSB signal occupies a smaller spectrum and is more spectrally efficient. A technique to overcome SSBI has been considered elsewhere \cite{sun2020experimental}.
In conclusion, our results show that the EVM for ARoF signals depends heavily on the allocated band, number of ROADMs and overall loss, suggesting that the system could be optimized depending on the target EVM. This could be carried out dynamically, considering real-time EVM measurements at the destination, as a result of both the fiber and radio transmission environments. 

%%%%%%%%%%%%%%%%%%%%%%%%%%%%%%%%%%%%%%%%%%%
\begin{figure}[h]
\centering
\includegraphics[width=1\linewidth]{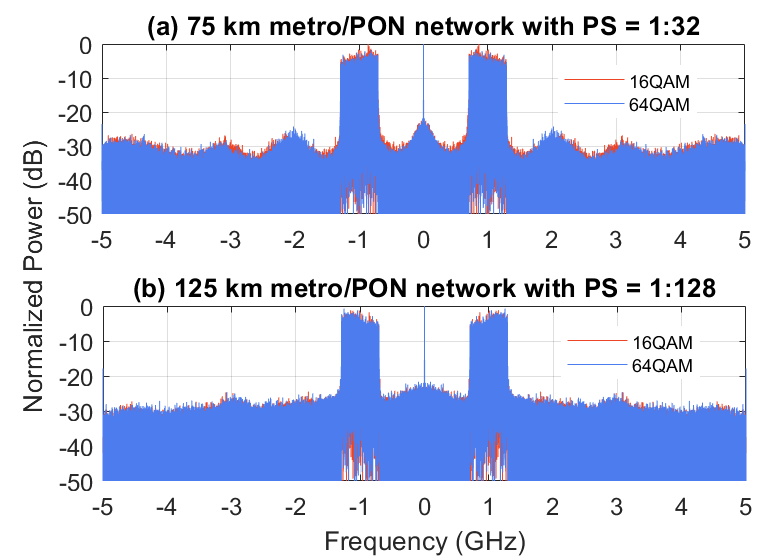}
\caption{Received electrical spectra of 16/64 QAM OFDM ARoF-1 signal for (a) a 75 km metro/PON network with 1:32 split and  (b) a 125 km metro/PON network with 1:128 split PON emulation.}
\label{fig:aclr}
\end{figure}
Another key metric for assessing transmission quality is the Adjacent Channel Leakage Ratio (ACLR), which quantifies the out-of-band power leakage from the carrier channel into adjacent frequency channels due to intermodulation effects. ACLR is determined by measuring the power difference between the carrier channel and adjacent channels. In an ARoF system, the degradation of SNR and ACLR is primarily influenced by several factors, including nonlinearities in the RF driver and amplifiers, the nonlinearity of optical-to-electrical and electrical-to-optical conversion processes, optical nonlinearities within the fiber, and scattering effects in the fiber \cite{lozhkin2015impact, 8552898}.
Figure \ref{fig:aclr} (a) and (b) illustrate the received electrical spectra of the 16/64 QAM OFDM ARoF-1 signals for two metro/PON topologies, with total fiber lengths of 75 km and 125 km, respectively, obtained by applying a fast Fourier transform to samples captured by the RTO. The out-of-band emissions are observed, and the ACLR is estimated between 23-30 dBc in Fig. \ref{fig:aclr} (a) and 23-24 dBc in Fig. \ref{fig:aclr} (b). In this work, the envelope detector is employed for downconversion from 60 GHz to the IF, which adds additional noise and degrades the SNR of the signal. Further optimization of the system's noise level, such as using low-noise electrical amplifiers, will help reduce ACLR. In addition, employing an electrical spectrum analyzer will improve the resolution bandwidth used in ACLR measurement.
%%%%%%%%%%%%%%%%%%%%%%%%%%%%%%%%%%%%%%%%%%%%%%
Thus, it is important to develop methods for quality of transmission estimation for ARoF signals, which is part of our future work.

\section{Conclusion}
This study investigates spectrally efficient techniques for heterogeneous transmission of DCO and mmWave ARoF signals across a converged metro and PON infrastructure. With the growing demand for high-density radio access deployments requiring high-capacity transmission at mmWave frequencies or beyond, upgrading legacy PON systems with advanced technologies to support both wireline and wireless services has become an attractive strategy for operators. This approach enables cost-efficient scaling while optimizing spectrum usage to tackle bottlenecks in metro networks.
%This work has demonstrated effective methods for optimizing hybrid signal transmission in converged metro and PON networks. 
In this work spectrum usage optimization was carried out by embedding ARoF signals within the existing DCO signal ROADM spectral window, to reduce the effect of coarse granularity (e.g., 6.25 GHz) of WSS bandwidth tunability (i.e., for ARoF signals which are of the order of a few GHz or less). We also showed how a unified design can connect metro nodes and PON endpoints using existing infrastructure, with new technologies added only at endpoints. Additionally, optical heterodyning with a carrier that can be shared over a PON was used for RF signal generation, simplifying endpoint design and lowering costs while supporting frequencies of 60 GHz and above, where there is more unoccupied spectrum. %These approaches show how next-generation metro-access networks can be made more efficient and cost-effective.
Further analysis of varying ROADM channel bandwidths and PON splitter ratios highlights their effects on signal integrity, emphasizing the importance of optimized design to meet performance targets such as EVM. This work provides actionable insights for deploying robust, cost-effective, and scalable converged metro/PON networks and opens the path to study QoT estimation methods for ARoF signals, to enable dynamic optimization of their EVM.

\section*{Acknowledgments}This work was supported by the European Commission project ECO-eNET, which has received funding from the Smart Networks and Services Joint Undertaking (SNS JU) under grant agreement No. 10113933. It was also co-funded by Research Ireland under grants 18/RI/5721 and 13/RC/2077 p2.

\bibliography{main}

\end{document}